\documentclass{aa}

\usepackage{graphicx}
\usepackage{txfonts}
\usepackage{enumitem}
\usepackage[breaklinks, colorlinks, citecolor=blue]{hyperref}
\usepackage{multirow}
\usepackage{natbib}
\usepackage{gensymb} 
\usepackage{ulem} 
\usepackage{color, soul} 

\begin{document}

   \title{Faraday tomography of LoTSS-DR2 data: II. Multi-tracer analysis in the high-latitude outer Galaxy}

   \titlerunning{Faraday tomography of LoTSS-DR2 data - draft}
   \authorrunning{Erceg et al.}
  
   \author{Ana Erceg\inst{1,2}$^\star$,
          Vibor Jelić\inst{1}$^{\star\star}$,
          Marijke Haverkorn\inst{2},
          Andrea Bracco\inst{3, 4},
          Lana Ceraj\inst{1},
          Luka Turić\inst{1},
          Juan D. Soler\inst{5}
          }

   \institute{Ruđer Bošković Institute, Bijenička cesta 54, 10 000 Zagreb, Croatia\\
              \email{$^\star$aerceg@irb.hr; $^{\star\star}$vibor@irb.hr}
         \and
             Department of Astrophysics/IMAPP, Radboud University, P.O. Box 9010, 6500 GL Nijmegen, The Netherlands 
        \and
             INAF–Osservatorio Astrofisico di Arcetri, Largo E. Fermi 5, 50125 Firenze, Italy 
        \and
            Laboratoire de Physique de l’Ecole Normale Supérieure, ENS, Université PSL, CNRS, Sorbonne Université, Université de Paris,
F-75005 Paris, France
        \and
            Istituto di Astrofisica e Planetologia Spaziali (IAPS). INAF. Via Fosso del Cavaliere 100, 00133 Roma, Italy           }
   \date{Received November 13, 2023; accepted April 10, 2024}

 
  \abstract
   {We conducted a follow-up study on the analysis of the LOw Frequency ARray (LOFAR) Two-metre Sky Survey (LoTSS) mosaic in the high-latitude outer Galaxy presented  in the first paper of this series. Here, we focus on the search for alignment between the magnetic field traced by dust, HI filaments, starlight optical linear polarisation, and linear depolarised structures (depolarisation canals) observed in low-frequency synchrotron polarisation. This alignment was previously found in several smaller fields observed with LOFAR, offering valuable insights into the nature of the interstellar medium and the 3D spatial distribution of the diffuse ionised medium.}
   {We aim to determine whether the alignment of the interstellar medium (ISM) phases observed through multiple tracers is a common occurrence or an exception. Additionally, in areas where depolarisation canals align with the magnetic field, we use starlight polarisation to constrain the distance to the structures associated with the observed canals.} 
   {We employed the Rolling Hough Transform (RHT) and projected Rayleigh statistics (PRS) to identify and quantify the alignment between the different tracers. We used these tools to detect linear features in the data and quantify the significance of the orientation trends between pairs of tracers. }
   {{On the scale of the whole mosaic, we did not find any evidence of a universal alignment among the three tracers. However, in one particular area, the western region (Dec between $29^\circ$ and $70^\circ$ and RA between $7\rm{^h}44\rm{^m}$ and $9\rm{^h}20\rm{^m}$), we do find a significant alignment between the magnetic field, depolarisation canals, and HI filaments. Based on this alignment, we used the starlight polarisation of stars with known parallax distances to estimate that the minimum distance to the structures observed by LOFAR in this region lies within the range of 200 to 240 pc. We associate these structures with the edge of the Local Bubble.}}

   \keywords{ Magnetic fields, polarisation, ISM: general, ISM: magnetic fields, ISM: structure, local interstellar matter, Radio continuum: ISM
               }

   \maketitle
%

\section{Introduction}
\label{Intro}
One of the largest projects of the LOw Frequency ARray \citep[LOFAR,][]{vanhaarlem13} project is LOFAR Two-metre Sky Survey \citep[LoTSS,][]{shimwell17, shimwell19, shimwell22}. Its aim is to observe the whole northern sky at low radio frequencies, from 120 to 168 MHz. Its scientific goals include investigating the formation and evolution of galaxies, black holes, as well as the  large-scale structures of the Universe and our Galaxy. The properties of the interstellar medium (ISM) and its magnetic fields were first studied on LoTSS preliminary data in the Hobby-Eberly Telescope Dark Energy Experiment \citep[HETDEX, ][]{hill08} spring field by \citet{vaneck19}. In our previous work \citep[][hereafter, E22]{erceg22} we expanded this research to a 5.5 times larger area using the second release of the LoTSS data \citep[LoTSS-DR2,][]{shimwell22}. There, we used the LoTSS-DR2 data to construct a 3\,100 squared-degree mosaic of synchrotron polarised emission and study the complexity of the magneto-ionic medium at low frequencies. In this work, we delve deeper by performing a multi-tracer analysis to better understand the low-frequency synchrotron polarised emission and its 3D distribution.

Linearly polarised synchrotron radiation emitted at a distance, $d,$ from the observer is affected by Faraday rotation as it passes through the magneto-ionic medium. The rotation changes the polarisation angle by $\Delta \theta$, which can be quantified through a parameter known as Faraday depth $\Phi$\footnote{In a special case of a background emitting source and foreground rotating medium, $\Phi$ can be called rotation measure \citep[RM, for the correct sense see][]{ferriere21}.}:

\begin{equation}\label{eq:FD}
     \frac{\Delta\theta}{\mathrm{[rad]}}=\frac{\lambda^2}{\mathrm{[m^2]}}\frac{\Phi}{\mathrm{[rad \ m^{-2}]}} = \frac{\lambda^2}{\mathrm{[m^2]}}0.81 \int_{0}^{d} \frac{n_e}{\mathrm{[cm^{-3}]}} \frac{B_\parallel}{\mathrm{[\mu G]}} \frac{dl}{\mathrm{[pc]}}.
\end{equation}
Here, $\lambda$ is the wavelength of observations, $n_e$ is the electron density, $B_\parallel$ is the component of the magnetic field along the line of sight (LOS), and the integral is taken over a path from the source to the observer at a distance, $d$. The Faraday depth has a positive or a negative sign depending on whether the magnetic field is orientated towards or away from the observer{, as per the standard convention used in Faraday studies} \citep[e.g.][]{ferriere21}. The degree of Faraday rotation is proportional to the wavelength's square, making this phenomenon more prominent at lower frequencies. This characteristic makes LOFAR exceptionally well suited for observing the effect of Faraday rotation. Using the technique of Faraday tomography, we can separate the polarised emission based on the amount of Faraday rotation it has experienced along the LOS \citep{burn66, brentjens05}. The result is a cube composed of polarised intensity images at different Faraday depths, namely, a Faraday cube. A Faraday spectrum at a specific position in the sky represents polarised intensity as a function of Faraday depth.

The Faraday cube of the E22 mosaic revealed that the observed area is filled with linear, depolarised structures known as the depolarisation canals \citep[see e.g. ][and references therein]{sokoloff98, haverkorn04, jelic15}. The depolarisation canals are elongated and narrow structures (akin to 'baguettes') of low or zero polarised intensity, formed by depolarisation at the scales of the telescope's resolution and defined with a point spread function (PSF). This can occur on a border of the two regions, where the polarisation angle abruptly changes by 90$^\circ$ {\citep[see an illustration in Fig. 2 of][]{fletcher&shukurov06}}. In that case, the border becomes depolarised and a canal with a width corresponding to the PSF size is formed. Canals formed in this way are frequency-dependent. Depolarisation canals can be frequency-independent when formed in a turbulent medium, where random polarisation angles cancel out within the scale of the PSF \citep{shukurov03}. While depolarisation canals are commonly observed in LOFAR observations \citep[][E22]{jelic15, jelic18, turic21}, the exact physical conditions that create them are not yet fully understood. Nevertheless, they can give us important clues in reconstructing the picture of the LOS distribution of {predominantly} ionised medium and the magnetic field permeating it.

\citet{jelic18} found an alignment between depolarisation canals, the orientation of HI filaments\footnote{The term HI filament in this paper refers to slender, linear features of HI found at the high Galactic latitudes defined as fibers in \citet{clark14}.} and the plane-of-sky magnetic field inferred by dust in a small field (5$^\circ$ $\times$ 5$^\circ$) centred on the 3C196 quasar. The authors claim that this result suggests that an ordered plane-of-sky magnetic field is crucial in confining different phases of the ISM, such as diffuse ionised gas and neutral components - HI gas and dust. \citet{bracco20} searched for alignment between structures observed in polarised synchrotron emission and different phases of the neutral medium and found a significant alignment in three additional fields surrounding the 3C196 field. The same three fields were studied in the work of \citet{turic21}, where an alignment between {\textit{Planck}} plane-of-sky magnetic field, depolarisation canals, and starlight polarisation was found. The authors were able to use this alignment to place initial constraints on the distance to the Faraday structures observed in one of the fields. This can be done in areas where the depolarisation canals align with the plane-of-sky magnetic field traced by linearly polarised dust emission and with the measurements of starlight polarisation angle. The key in this alignment is the magnetic field, which aligns the dust grains; these, in turn, emit partially linearly polarised sub-millimeter emission and polarise the light of the background stars in the direction of the magnetic field \citet{davis&greenstein51, lazarian07, andersson15}. { While the mechanism by which the depolarisation canals may become aligned with this magnetic field is not yet understood, an alignment between these tracers is observed in some sky areas, which indicates a causal connection. If the magnetic field is responsible for aligning both depolarisation canals and starlight polarisation, we can imagine several 3D scenarios that could produce an alignment in projection to the plane of sky. The magnetic field could be localised and aligning different tracers at one particular distance -- or it could be uniform across larger volumes of space, aligning tracers at various distances from us. There is no way to differentiate between the two configurations using the aforementioned tracers. Therefore, by searching for alignment and using the distances to the background stars, we may determine only the minimum distance to the Faraday structures associated with the depolarisation canals.}

The distance \citet{turic21} found to the structures associated with depolarisation canals was around 200 pc, which puts it in our nearest Galactic neighbourhood. This result aligns with the idea that LOFAR's high sensitivity to Faraday rotation makes it susceptible to signal depolarisation over long distances. It is therefore expected that LOFAR is mostly probing the local Galactic medium, such as the nearby radio Loop III \citep{quigley65, berkhuijsen1971a} observed in E22, or possibly the wall of the Local Bubble \citep[][]{cox87}. Local Bubble is an irregular cavity in the Galactic ISM that encapsulates the Solar system. Its edge can be found at distances of several hundreds of parsecs, depending on the direction of the observation \citep{lallement03}. Recent dust extinction maps \citep{lallement19} have enabled the construction of a 3D model of the Local Bubble shell by \citet{pelgrims20}. The modelled distances to the inner wall of the Local Bubble can be compared with distance estimates obtained from our analysis.

This paper aims to advance our understanding of the 3D distribution of the Galactic magneto-ionic medium through multi-tracer analysis. By analysing the relation between different tracers of the ISM -- HI filaments, depolarisation canals, and magnetic field -- we aim to understand whether the alignment between them {on the plane of sky} is universal across large areas. Furthermore, we aim to utilise the alignment between depolarisation canals, magnetic field and starlight polarisation to infer the distance to the structures observed in LoTSS-DR2 mosaic of E22.

This paper is organised as follows. In Sect. \ref{sec:data}, we describe the datasets used in this paper. In Sect. \ref{sec:stat_tools}, we introduce the main statistical tools used in our analyses:\ the Rolling Hough Transform (RHT) and projected Rayleigh statistics (PRS). We present the results of analyses of the orientation trends between different tracers in Sect. \ref{sec:results} and comment on their implications in Sect. \ref{sec:discussion}. We present our conclusions in Sect. \ref{sec:conclusions}.

\section{Data}
\label{sec:data}
In this section, we describe the datasets used in the paper. All the datasets are constrained to the sky area of the LoTSS mosaic (E22), from $\mathrm{8\rm{^h}00\rm{^m}}$ to $\mathrm{16\rm{^h}40\rm{^m}}$ in RA and from $\mathrm{29^\circ}$ to $\mathrm{70^\circ}$ in Dec. The LoTSS data are described in Sect. \ref{sec: lotssdata}, followed by the description of the auxiliary data used to examine the alignment of different ISM tracers and define the distance of the structures seen in the LoTSS mosaic. 

\subsection{LoTSS data}
\label{sec: lotssdata}
Throughout the paper, we analyse the maximum polarised intensity image created using the LoTSS mosaic Faraday cube\footnote{Faraday cube and derived products described in \citet{erceg22} are publicly available at FULIR database \url{https://data.fulir.irb.hr/islandora/object/irb:367}.}, both presented in E22. The cube was constructed using low-frequency LoTSS-DR2 observations covering a frequency range from 120 to 168 MHz, which is now publicly available. The Faraday tomography was performed using the RM-synthesis algorithm \citep{brentjens05}, which utilises a modification of a Fourier transform to convert frequency-dependent spectra into Faraday depth-dependent spectra of polarised intensity, presented in a single Faraday cube. {The primary advantage of using RM-synthesis is its capacity to reduce noise by $\sqrt{N}$, with N representing the number of input frequencies. Low-intensity structures observed by LOFAR are challenging to distinguish from noise in individual $Q$ and $U$ images and this reduction enables us to observe them in the final Faraday cube.} The mosaicked Faraday cube has an angular resolution of 5.5 arcmin and covers the Faraday depth range from -50 to +50 $\mathrm{rad \ m^{-2}}$, with the resolution in Faraday depth being 1 $\mathrm{rad \ m^{-2}}$. The noise level in the cube is $71~{\rm \mu Jy~PSF^{-1}~RMSF^{-1}}$. The three-dimensional (3D) Faraday cube was collapsed into a maximum polarised intensity map by selecting the peak value in the Faraday spectrum for each LOS. The resulting image reveals the presence of depolarisation canals. To identify and quantify the orientation of the canals, we employed the RHT method, as detailed in Section \ref{sec:stat_tools}.

\subsection{Starlight polarisation data}
\label{star_pol_data}
We used starlight polarisation measurements in combination with distances to the stars to set the minimum limit to the distance of Faraday structures associated with depolarisation canals. We searched for catalogues containing measurements of starlight polarisation angle in the \texttt{VizieR}\footnote{\url{https://vizier.cds.unistra.fr/viz-bin/VizieR}} database and used data from the same catalogues as \citet{turic21} \citep{heiles20, berdyugin01, berdyugin02, bailey10, berdyugin14}, with the addition of \citet{piirola20}. We obtained the distances to these stars by cross-matching the sources from these catalogues with the \citet{bailerjones21} catalogue based on the Gaia Early Data Release 3 \citep{gaiaedr3}. 

Some of the stars had multiple polarisation measurements available from different catalogues; in such cases, we selected the measurement with the lowest error in starlight polarisation angle. Furthermore, we excluded all the stars with starlight polarisation angle errors larger than 60 degrees. After applying these filters, we were left with 709 stars in total.

\subsection{\textit{Planck} magnetic field data}
The plane-of-sky magnetic field as traced by dust can be obtained from \textit{Planck} maps of Stokes parameters Q and U\footnote{Publicly available at The Planck Legacy Archive: \url{https://pla.esac.esa.int}} at 353 GHz \citep{planck20}. The plane-of-sky magnetic field orientation is calculated by rotating the dust polarisation angle from the maps by 90 degrees. To increase the signal-to-noise ratio (S/N), we smoothed the data to a resolution of 1 degree.

\subsection{EBHIS data}
We used the 21 cm spectroscopic HI data from the Effelsberg-Bonn HI survey\footnote{\url{http://cdsarc.u-strasbg.fr/viz-bin/qcat?J/A+A/585/A41}} \citep[EBHIS, ][]{winkel16} to search for alignment between the depolarisation canals and the neutral phase of the ISM on the large scale of LoTSS mosaic.

The EBHIS data come in the form of position-position-velocity (PPV) cubes with an angular resolution of 10.8 arcmin and a spectral resolution of $1.44~\mathrm{km~s^{-1}}$. The full data cover the local-standard-of-rest velocity range of $|v_{LSR}| < 600 \mathrm{km~s^{-1}}$, however, significant levels of emission in the area of LoTSS mosaic are found in velocity range from $-139.2$ to $+108.5~\mathrm{km~s^{-1}}$. As we expect the distance to the structures analysed in this paper to be at most at few hundred parsecs from the Sun, we avoided velocity ranges occupied by intermediate to high velocity clouds likely coming from cold gas in the distant halo \citep[farther than 1 kpc, e.g.][ and references therein]{marasco22}. For this reason, we integrated the HI intensity over the range of $|v_{LSR}| < 30 \mathrm{km~s^{-1}}$.

\section{Methods and statistical tools}
\label{sec:stat_tools}
For the robust statistical analysis of the alignment between different tracers, we used the following statistical tools.

\subsection{Rolling Hough Transform}
The RHT \citep[][]{clark14} is a computer vision and image processing algorithm based on the Hough algorithm \citep{hough62}, used to identify straight lines in an image. It can be used in the analysis of linear features of the ISM, such as HI filaments \citep{clark14, clark19} or depolarisation canals \citep{jelic18, turic21}. The algorithm allows us to quantify linear features in an image by assessing the probability that each pixel is a part of a coherent linear structure oriented in a specific direction. In this paper, we use it to define and quantify linear structures in both LOFAR and EBHIS datasets.

The output of the RHT, $R(\theta, x, y)$, represents the probability that a pixel located at $(x, y)$ in the image is part of a coherent straight line with a direction angle $\theta$. The angle $\theta$ is defined in the range of [$0^\circ$, $180^\circ$], where $0^\circ$ points to the north of the image. There are three input parameters of RHT that can be tuned to enhance specific linear features in the image. The probability threshold ($Z$) defines the lowest probability $R(\theta, x, y)$ that is considered acceptable. The smoothing kernel diameter ($D_K$) specifies the size of a two-dimensional kernel used to smooth and unsharp mask the data, resulting in sharpened lines and reduced large scale features. The window diameter ($D_W$) defines the diameter of a circular window within which the algorithm searches for coherent linear features; therefore, it defines the shortest line RHT can detect. \citet{jelic18} demonstrated that detection of depolarisation canals using RHT is robust to changes in the parameters. Modifying the parameters of the RHT only impacts the level of noise in the data, with the mean directions of detected features remaining unchanged.

To produce a distribution of the orientation of structures in a specific area, we integrated the RHT distributions $R(\theta, x, y)$ over that area, as follows: 

\begin{equation}
    \Tilde{R}(\theta) = \int \int_{area} R(\theta, x, y) \ dxdy,
\end{equation}
per the definition in \citet{clark14}. From this distribution, we can calculate the mean, $\langle \theta \rangle$, and the spread of the distribution, $\delta\theta$, following \citet{jelic18}. By projecting angle $\theta$ to a full circle and defining every point as a vector of length $\Tilde{R}^2(\theta) d\theta$, they defined a vector of S as:

\begin{equation}
    S = \frac{\int_{-\pi/2}^{\pi/2} \Tilde{R}^2(\theta)e^{2i\theta}d\theta}{\int_{-\pi/2}^{\pi/2} \Tilde{R}^2(\theta)d\theta}.
\end{equation}
 The direction and length of $S$ can be used to measure $\langle \theta \rangle$ and $\delta\theta$:

\begin{equation}
     \langle \theta \rangle = \frac{1}{2}\rm{arctan2}\frac{\rm{Im}(S)}{\rm{Re}(S)},
\end{equation}

\begin{equation}
     \delta\theta = \frac{1}{2}\sqrt{\rm{ln}(1/|S|^2)}.
\end{equation}

The output of RHT can be visualised by creating a backprojection, namely, by integrating $R(\theta, x, y)$ over all the angles $\theta$. In this case, the intensity of pixels reflects the probability that they belong to a coherent linear structure. 

\subsection{Projected Rayleigh statistics}
An optimal statistics that can be used to quantify the alignment between two reference orientations of $0^\circ$ or $90^\circ$ is the PRS \citep[,][]{jow18}. We used it to test the hypothesis that the distribution of angle differences between two tracers peaks around zero, thus indicating that the two tracers are aligned.

We followed the procedure from \citet{panopoulou21} to calculate the PRS, estimate its uncertainty, and test whether the alignment is statistically significant. \citet{panopoulou21} defines PRS as:

\begin{equation}
    PRS = \frac{1}{\sqrt{\sum_i^N\frac{w_i^2}{2}}}\sum_i^Nw_i \cos2\Delta\psi_i.
\end{equation}
Here we define the angle difference $\Delta\psi$ between two angles $\theta_1$ and $\theta_2$ as:
\begin{equation}\label{eq:deltapsi}
    \Delta \psi = \frac{1}{2} \arctan \left( \frac{\sin2\theta_1 \cos2\theta_2-\cos2\theta_1 \sin 2\theta_2}{\cos2\theta_1\cos2\theta_2+\sin2\theta_1\sin2\theta_2}\right),
\end{equation}
to account for $\pi$-ambiguity \citep{planck16c}. We define the weights, $w_i$,
\begin{equation}
    w_i = \frac{1}{\sigma_{\Delta \psi}^2},
\end{equation}
where $\sigma_{\Delta \psi}$ accounts for uncertainties in both angle measurements, $\sigma_{\theta_1}$ and $\sigma_{\theta_2}$. Since the uncertainties in RHT and starlight polarisation angle are not necessarily Gaussian, we use a Monte Carlo (MC) method to estimate $\sigma_{\Delta \psi}$. Positive PRS values indicate the two orientations are parallel, while negative PRS values indicate perpendicular relative orientation. 

The statistical significance of our results is determined by running a MC simulation for a set of angles coming from normal distributions whose mean values are uniformly distributed in the range $[-90^\circ, 90^\circ]$. At the same time, standard deviations are the fundamental observed uncertainties. The simulated data represent a set of angle differences between random angles measured with the same precision as our dataset. Supposing the PRS of the observed distribution is greater than the 99.9th percentile of the PRS distribution coming from MC simulations, it is statistically unlikely (at a level of $3 \sigma$ or more) that the observed distribution is random; thus, we would deem the alignment as statistically significant. We call this value the significance threshold. To estimate the PRS uncertainties, we again used an MC method, drawing a sample of angles from normal distributions with mean values equal to $\Delta \psi$ values and standard deviations equal to values of $\sigma_{\Delta \psi}$. We calculated the PRS distribution for 10 000 sets of angle samples. The standard deviation of the obtained PRS distribution corresponds to the PRS uncertainty. 

\section{Analysis and results}
\label{sec:results}
In the following subsections, we apply the statistical tools outlined in Sect. \ref{sec:stat_tools} to assess the alignment between depolarisation canals and three other observational tracers:\ starlight polarisation, magnetic field, and HI filaments. We study the alignment between the tracers only in projection, which does not necessarily imply existence of alignment in 3D space, where the morphology may be more complex. A complete 3D analysis calls for realistic large scale simulations of the multi-phased ISM, which are beyond the scope of this paper.

We use RHT to find and quantify linear features in LoTSS and HI data images and PRS to examine whether a pair of tracers shows alignment. First, we present the results of the RHT algorithm performed on the LoTSS mosaic.

\subsection{Depolarisation canals}

\begin{figure*}
   \centering
   \includegraphics[width=\hsize]{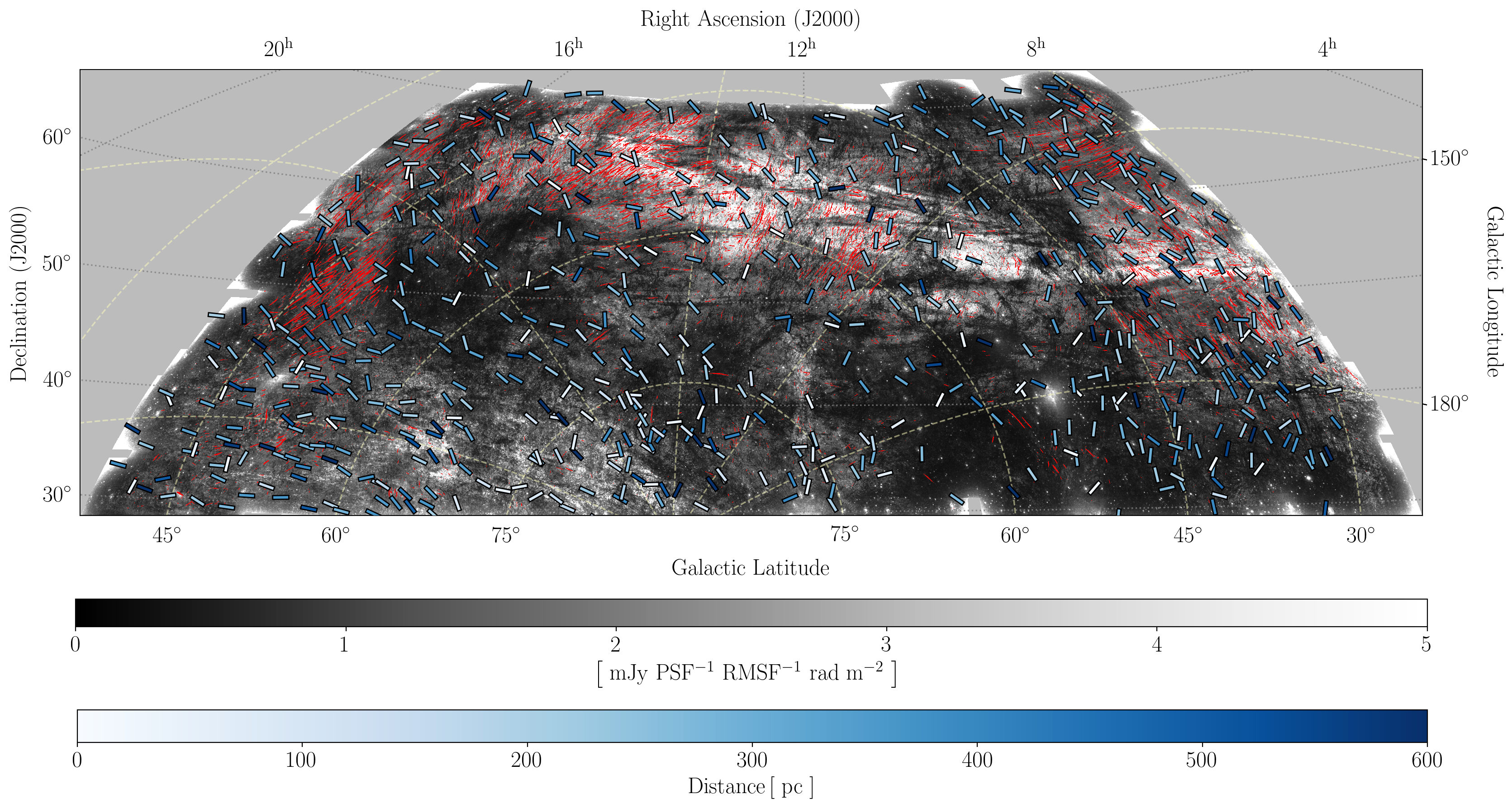}
      \caption{RHT backprojection depicting depolarisation canals (red lines) overplotted on the maximum polarised intensity image of the whole LoTSS mosaic. The blue bars on top of the two represent the sample of starlight polarisation measurements. They are coloured according to their distance, as indicated by the lower colour bar. The orientation of the lines represents the orientation of the starlight polarisation angle in the same projection as the LoTSS data.}  \label{RHT+stars}
\end{figure*}

\begin{figure}
   \centering
   \includegraphics[width=\hsize]{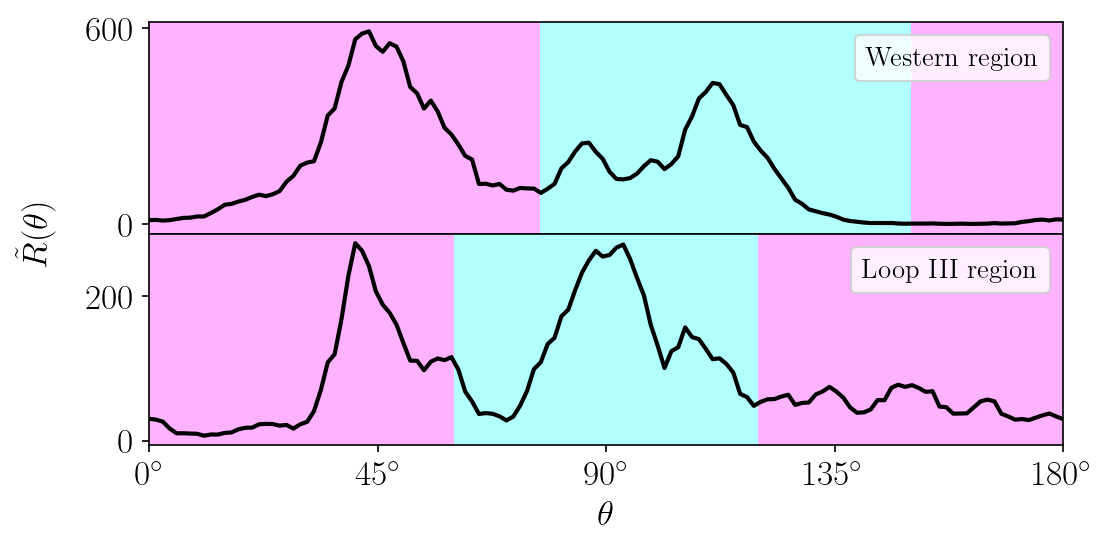}
      \caption{Distribution of the orientation of depolarisation canals in the western region, shown in the upper panel. Angles highlighted in cyan and magenta represent the east-west and the north-south orientation, respectively. The distribution of the orientation of depolarisation canals in the Loop III region is shown in the lower panel. Angles highlighted in cyan and magenta represent the depolarisation canals perpendicular and parallel to the loop, respectively. }   \label{RHT_spectra}
\end{figure}

\begin{figure*}
   \centering
   \includegraphics[width=\hsize]{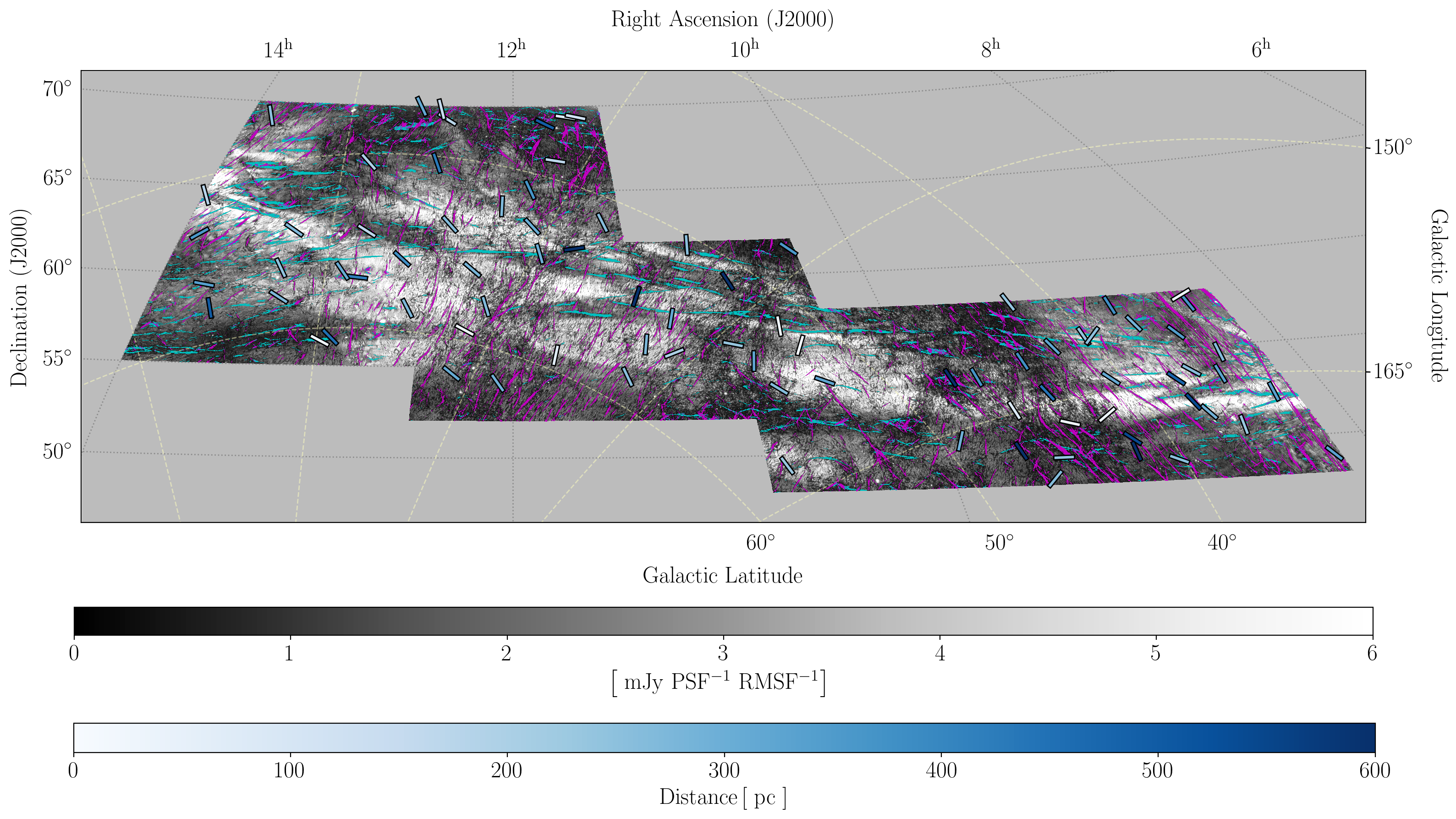}
      \caption{Zoom on the maximum polarised intensity image in the Loop III region. Cyan and {magenta} lines represent the depolarisation canals perpendicular and parallel to the loop, respectively. Blue bars represent the orientation of starlight polarisation angles, while the distance to the stars is indicated by the lower colour bar.}   \label{loop_RHT}
\end{figure*}

\begin{figure*}
   \centering
   \includegraphics[width=\hsize]{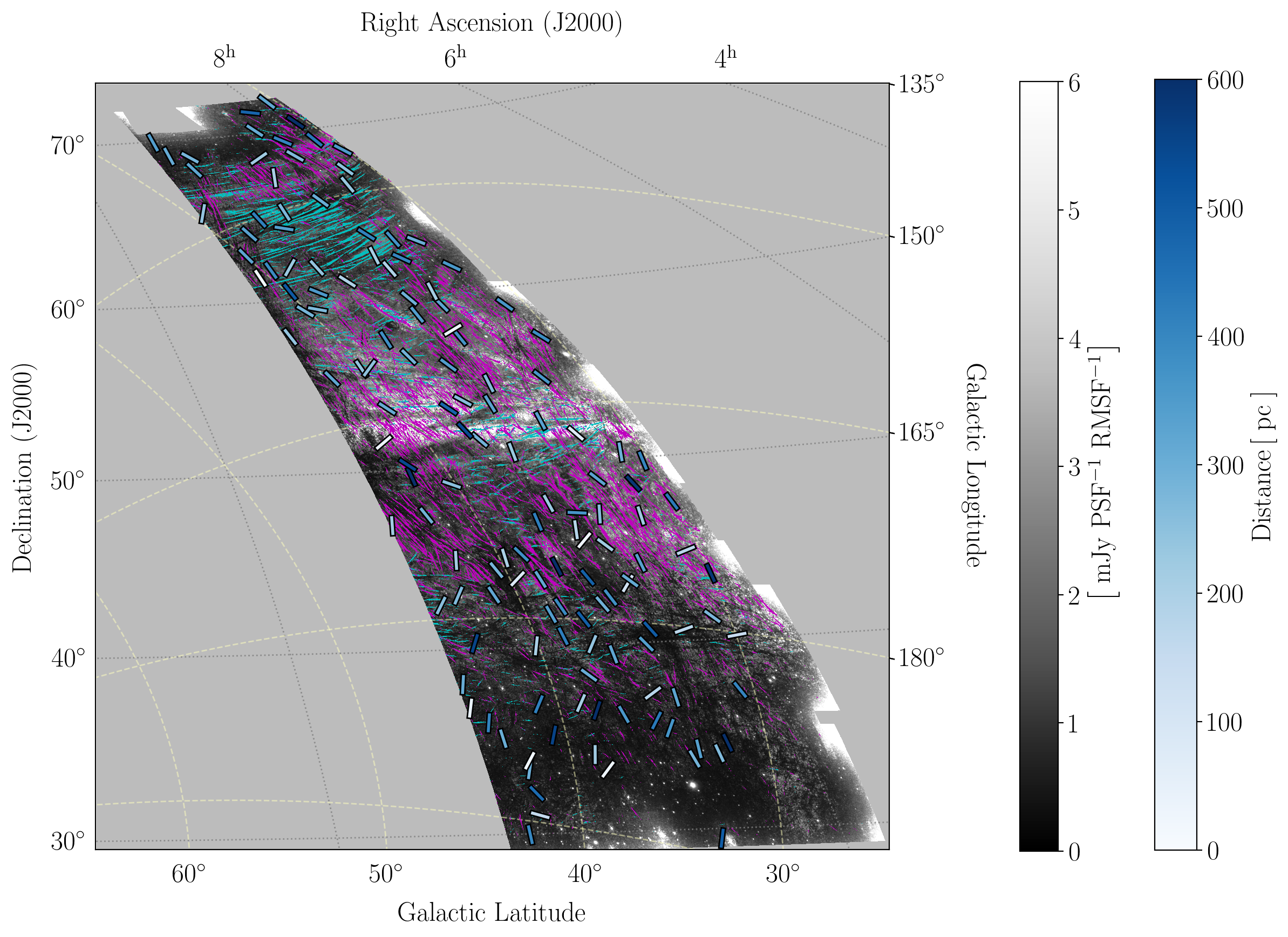}
      \caption{Zoom on the maximum polarised intensity image in the western region. Cyan and {magenta} lines represent the depolarisation canals in the east-west and north-south direction, respectively. Blue bars represent the orientation of starlight polarisation angles, while the distance to the stars is indicated by the colour bar on the right.}   \label{west_RHT}
\end{figure*}

By inverting the image of maximum polarised intensity through calculation of the reciprocal, the depolarisation canals are enhanced while the surrounding emission is suppressed. We applied RHT to the inverted images of maximum polarised intensity using the following RHT parameters: ($D_K$, $D_W$, $Z$) = (10, 61, 0.8), which corresponds to ($D_K$, $D_W$, $Z$) = (8 arcmin, 50 arcmin, 0.8) at LoTSS resolution. Choice of the parameters is the same as in \citet{jelic18} and \citet{turic21}. The RHT backprojection is presented in Fig. \ref{RHT+stars}, superimposed on the image of maximum polarised intensity. 

At {declinations below $\sim45\degree$}, which roughly correspond to higher Galactic latitudes, the emission is fainter, patchy, or noise-dominated. In these regions, depolarisation canals are sparse or absent and the short and randomly oriented lines detected by the RHT algorithm represent features created by the noise. Noise features have a low probability of being a part of a linear structure, which is reflected in their low RHT intensity; this, in turn, minimises their impact on the results of our analysis.

We have identified well-defined depolarisation canals in two regions of interest. The first region is associated with {emission connected to Loop III}. It is approximated by {a union of} three rectangles, as shown in a cutout of the LoTSS mosaic in Fig. \ref{loop_RHT}. {The first rectangle spans the Dec range from $55\degree$ to $70\degree$ and RA range from $11\rm{^h}20\rm{^m}$ to $14\rm{^h}00\rm{^m}$, the second one spans over Dec range from $52\degree$ to $62\degree$ and RA range from $10\rm{^h}20\rm{^m}$ to $12\rm{^h}30\rm{^m}$, while the third one spans the range from $48\degree$ to $50\degree$ in Dec and from $8\rm{^h}10\rm{^m}$ to $10\rm{^h}50\rm{^m}$ in RA.} This definition of the Loop III region is more restricted compared to the one in E22, as we aimed to include only the areas with high intensity of diffuse emission and well-defined depolarisation canals. The second region of interest is the area of ordered depolarisation canals in the west, covering Dec between $29^\circ$ and $70^\circ$, and RA between $7\rm{^h}44\rm{^m}$ and $9\rm{^h}20\rm{^m}$, as depicted in Fig. \ref{west_RHT}. This region includes the 3C196B field analysed in \citet{bracco20} and \citet{turic21}, centred at RA, Dec = $8\rm{^h}13\rm{^m}$, $+40^\circ 24'$. 

The distribution of the orientation of depolarisation canals in the two regions is shown in Fig. \ref{RHT_spectra}. Similarly to the canals observed in the 3C196B field, the depolarisation canals in both regions are distributed in two main orientations. The work of \citet{turic21} suggest that the structures associated with each group of these depolarisation canals may be located at different distances along the LOS. Therefore, we want to search for alignment between either of the depolarisation canals groups and the other tracers. 

In each of the studied regions, the depolarisation canals often intersect with each other, which is why it was necessary to bin them by their orientation and perform the analysis separately for each group. In the Loop III region, we categorise depolarisation canals into two groups: those that follow the shape of the loop itself and those that are perpendicular to the loop (as indicated by cyan and magenta lines in Fig. \ref{loop_RHT}). The ones following the loop have inclinations relative to the north of the image ranging from $60^\circ$ to $120^\circ,$ while the ones perpendicular to the loop have inclinations ranging from $120^\circ$ to $60^\circ$. In the western region, we distinguish between depolarisation canals oriented east to west (ranging from $77^\circ$ to $150^\circ$) and those oriented north to south (ranging from $150^\circ$ to $77^\circ$, see cyan and {magenta} lines in Fig. \ref{west_RHT}). The former ones are mostly found at declinations above $50^\circ$, while most of the latter ones fill the declination range from $\sim35^\circ$ to $70^\circ$. Our selection of bins is depicted in Fig. \ref{RHT_spectra}. Slight changes in the binning do not change the results significantly, since each depolarisation canal is weighted by its RHT intensity (probability) throughout the analysis.

The width of the depolarisation canals in LoTSS mosaic matches the size of the PSF, while the average length is approximately on the order of $1^\circ$. The longest depolarisation canals stretch over up to $20^\circ$ and are found in the westernmost and easternmost areas of the mosaic. Precise statistics of the depolarisation canals lengths requires a more accurate definition of the borders of a single depolarisation canal. This aspect is beyond the scope of this paper, but will be addressed in future work.

\subsection{\textit{Planck} magnetic field}
\label{Sect:mag field vs lotss}
\begin{figure*}
   \centering
   \includegraphics[width=\hsize]{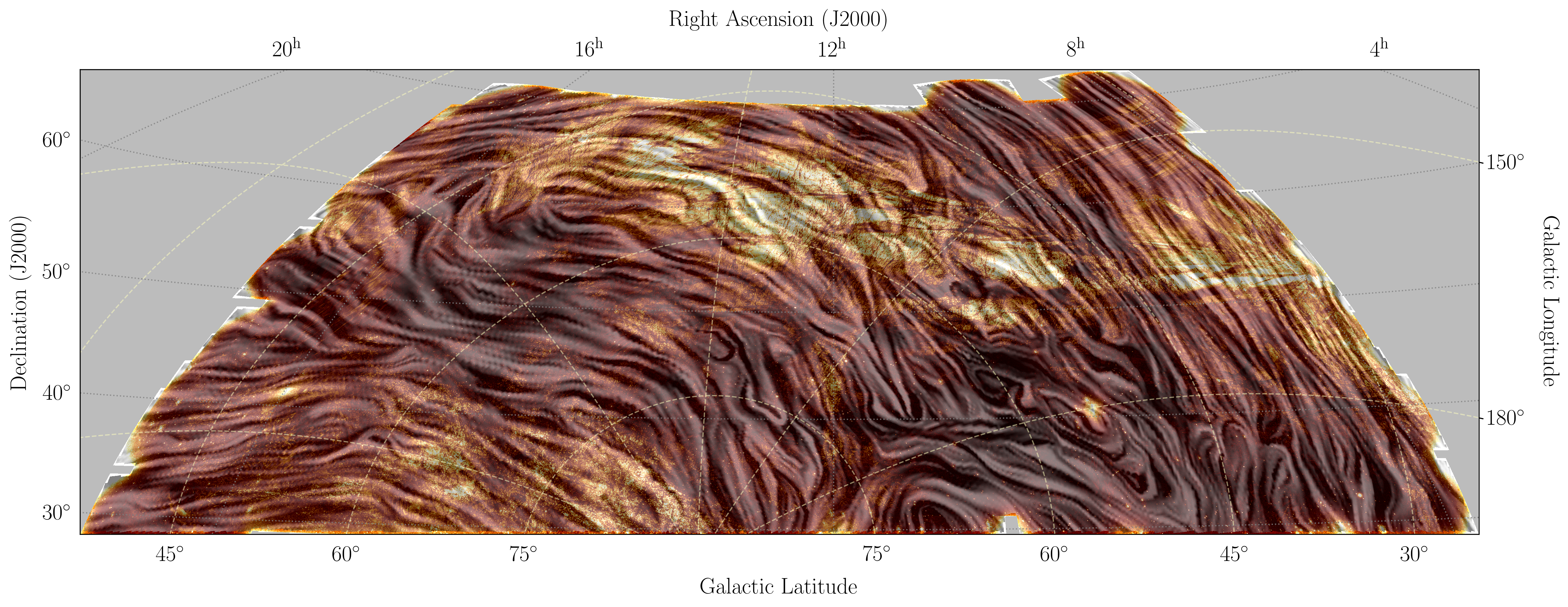}
      \caption{Magnetic field orientation from \textit{Planck} at 353 GHz, overplotted on the image of maximum polarised intensity of the LoTSS mosaic. }   \label{mag_field}
\end{figure*}

\begin{figure}
   \centering
   \includegraphics[width=\hsize]{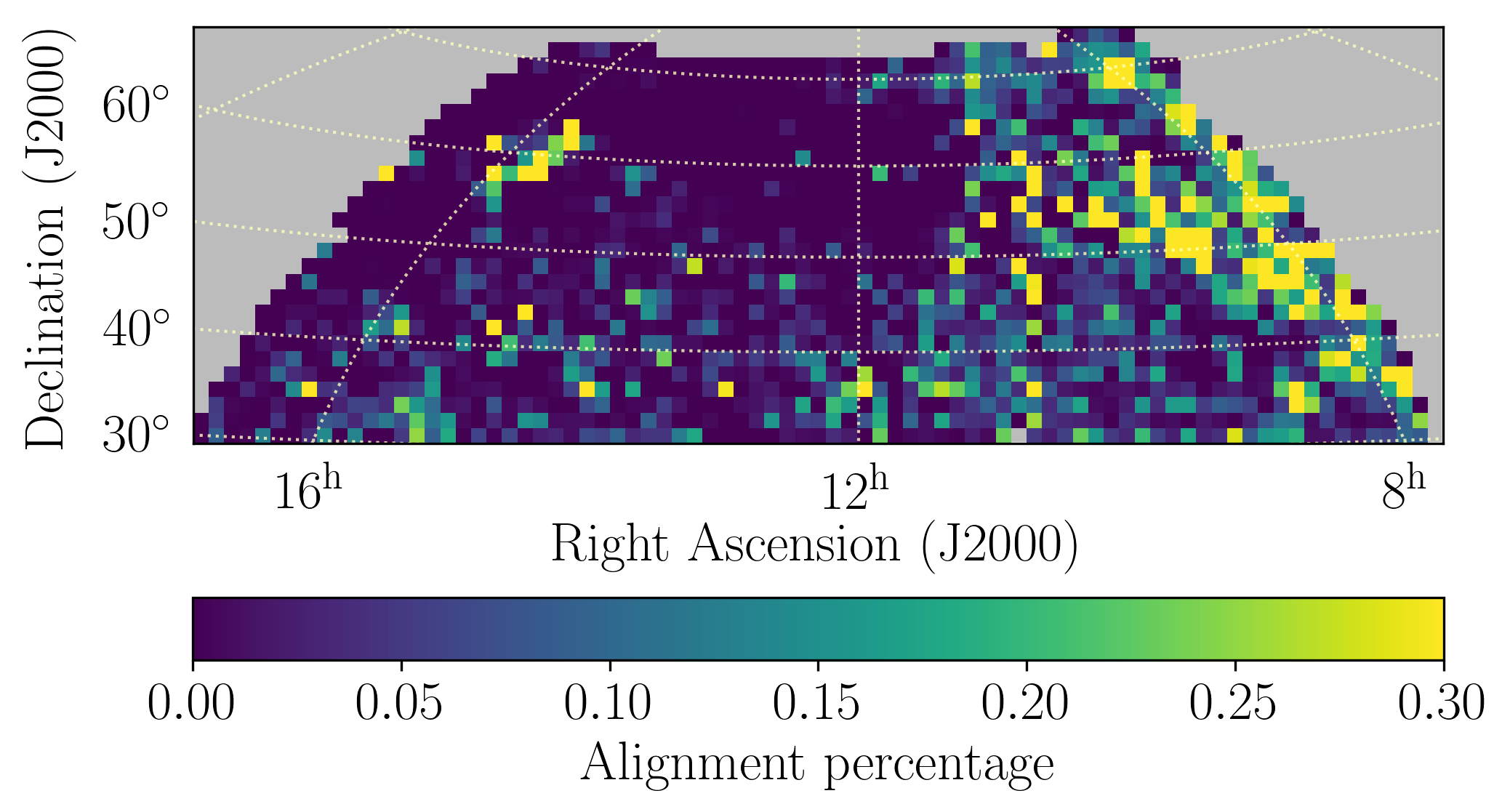}
      \caption{Ratio of depolarisation canals that align with the \textit{Planck} plane-of-sky magnetic field.}   \label{mag_field_alignment}
\end{figure}

\begin{table}
 \caption[]{\label{tab_magfield_PRS} Results of the PRS analysis for the alignment between the direction of the plane-of-sky magnetic field {from \textit{Planck}} and the depolarisation canals. Results indicating significant alignment are written in boldface.}
\begin{tabular}{lcccc}
 \hline \hline
  Region &
  Orientation bin &
  PRS &
  $\sigma_\mathrm{PRS}$ &
  significance 
\\
 &
  &
   &
   &
  threshold
 \\ \hline

Loop III    &   $60^\circ-120^\circ$  & -4.0 & 0.2 & 2.7\\
    &   $120^\circ-60^\circ$  & 1.4  & 0.4 & 2.8\\
    \hline
Western        &   $77^\circ-150^\circ$  & -5.9 & 0.2 & 2.8\\
region        &   $150^\circ-77^\circ$  & \bf{8.8} & \bf{0.5} & \bf{3.0} \\

\hline
\end{tabular}

\end{table}

We utilised \textit{Planck} data to reconstruct the plane-of-sky magnetic field in the area of the mosaic, and present the results in Fig. \ref{mag_field}. We calculated the PRS for the same two regions of interest as in the previous section. To create a sample of angles for PRS, we divide each region into square patches with sides of 1.5 degrees. In each patch, we calculated the circular mean of the the magnetic field orientation, $\theta_\mathrm{mag}$, and the circular mean of the depolarisation canals orientation $\theta_\mathrm{DC}$. The difference between these two orientations gives us the $\Delta \psi$ of the PRS statistics (see Eq. \ref{eq:deltapsi}). We report the PRS values, their errors, and the significance thresholds in Table \ref{tab_magfield_PRS}. Negative PRS values in the table, indicate that the chosen bin of depolarisation canals is perpendicular to the magnetic field. This means that the { \textit{Planck} magnetic field is not dominated by dust emission that occupies the same volume of space as the structures associated with the depolarisation canals, which prevents us from using} starlight polarisation to determine the location of {these} structures.

The magnetic field in the western region is ordered and oriented in the north-south direction, aligning with the north-south depolarisation canals. We found this alignment to be statistically significant (see Table \ref{tab_magfield_PRS}). As a consequence, a negative PRS value is obtained for comparison with the depolarisation canals in the east-west direction.

{In the Loop III} region, the magnetic field does not exhibit morphological correlation with the shape of the loop, which indicates that Loop III is not the dominant source of dust along these LOS. The amount of dust in the loop seems insufficient to induce significant stellar polarisation that would accurately trace the plane-of-sky magnetic field of the loop. Therefore we cannot differentiate between stars in front and behind the loop and determine the distance to the loop. 

In Fig. \ref{mag_field_alignment}, we present a map of alignment ratio between the orientation of the {\textit{Planck}} plane-of-sky magnetic field and the depolarisation canals. We used the same patches as in the PRS analysis. For each patch, we calculated the mean magnetic field and counted all the depolarisation canals whose orientation falls within 10 degrees of that mean value. The alignment ratio represents the number of pixels containing depolarisation canals that are aligned with the magnetic field, divided by the total number of pixels containing a depolarisation canal. The highest alignment ratio is found in the western region. 

\subsection{Depolarisation canals and starlight polarisation}

\label{stellar pol vs dep canals}

\begin{figure}
   \centering
   \includegraphics[width=\hsize]{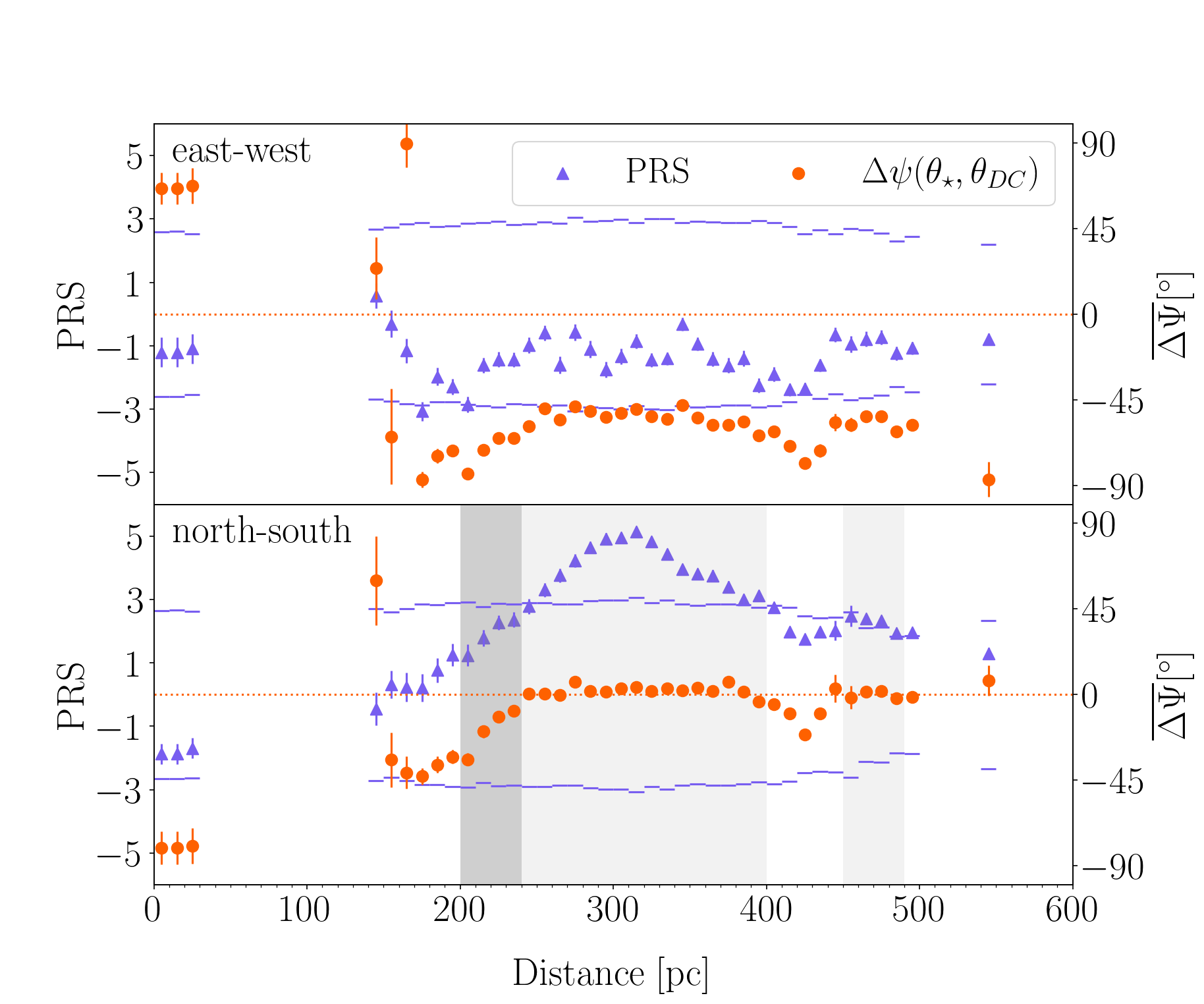}
      \caption{ PRS values (purple triangles), their errors and the significance threshold (short, horizontal bars) calculated for angles $\Delta \psi$, between starlight polarisation and depolarisation canals in the western region. Additionally, the $\Delta \psi$ (orange circles) and the associated error are shown. The result for each bin is represented by a data point located at its lower border. The two panels represent two distinct sets of orientations of depolarisation canals, as noted in the top left corner of each panel. Lighter shaded area represents all bins in which we find alignment, while the darker shaded area marks the estimated minimum distance range. }   \label{west_PRS}
\end{figure}

\begin{figure}
   \centering
   \includegraphics[width=\hsize]{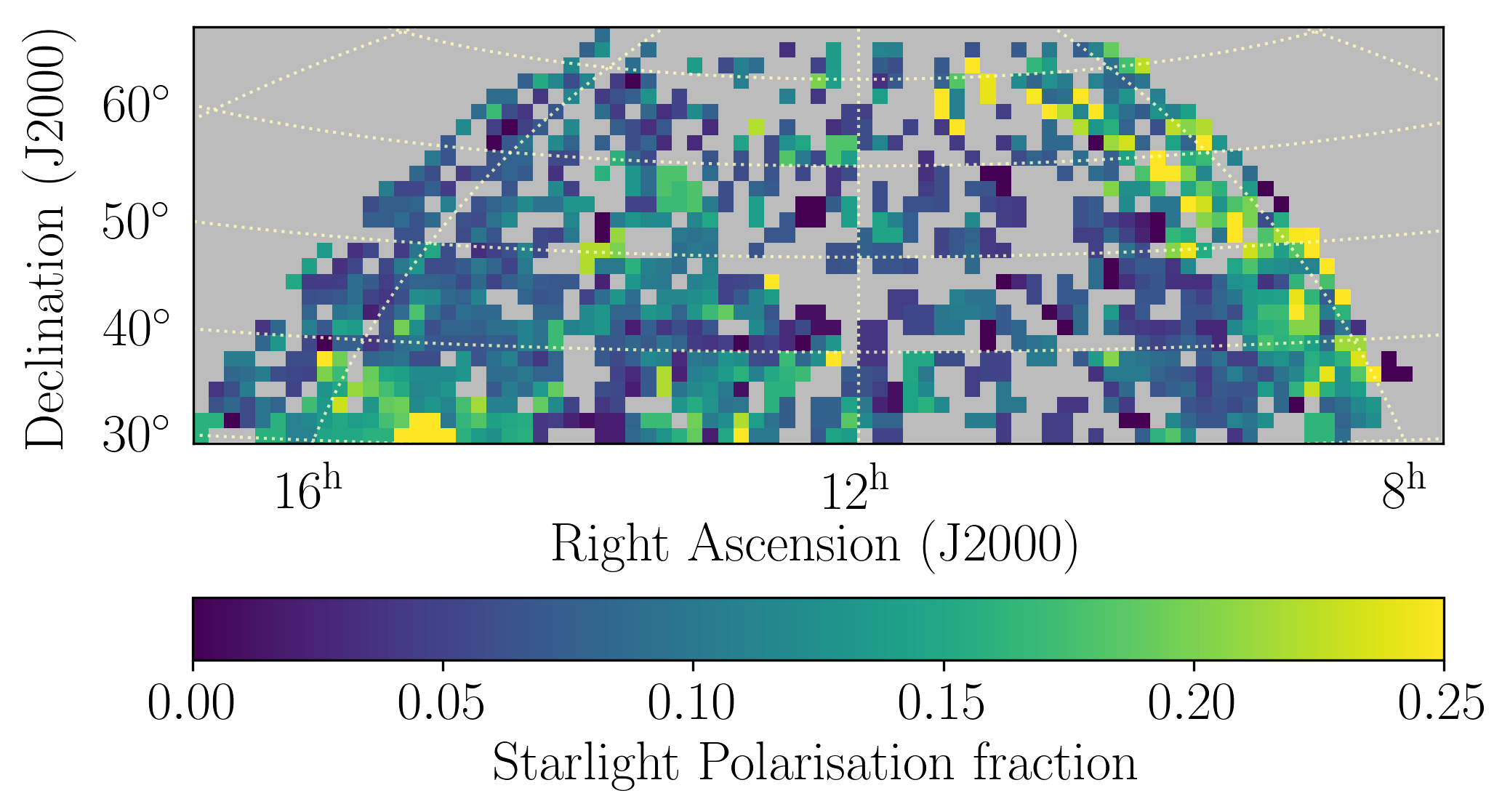}
      \caption{Starlight polarisation fraction. Each patch has a side of $1.5^\circ$ and its value represents the average polarisation fraction of all stars found in the patch.}   \label{pol_frac}
\end{figure}

The significant alignment between the depolarisation canals and the magnetic field orientation in the western region of the mosaic indicates that the two are physically connected. We assume that they are located at the same distance from the Sun. Thus, we proceeded with the search for alignment between depolarisation canals and the starlight polarisation in that region.

We can estimate the distance to the structures associated with depolarisation canals by finding the distance at which starlight polarisation angles begin to align with the depolarisation canals. To do so, we divided a subset of stars in the western region (159) into distance bins. The number of measurements in each bin varies, due to differences in stellar measurement density. If a bin has fewer than five measurements, we excluded it from the analysis. To mitigate the effect of choosing the first bin location on the distance estimate, we shifted the bin locations by a step size smaller than the bin size, following the approach used in \citet{panopoulou21}. For example, with a bin size of 100 pc and a step size of 10 pc, the first three bins would be: [0, 100] pc, [10, 110] pc, and [20, 120] pc. This procedure effectively duplicates the data that we have and makes the final plot useful only in the context of determining the lower and the upper limits for the distance to the structure.   

The local orientation of depolarisation canals is determined by calculating the mean and the standard deviation of RHT within a circular region centred on each star, with a radius of 75 pixels ($\sim 1^\circ$). At a distance of 200 pc, this corresponds to a radius of roughly 7 pc. The angle $\Delta \psi$ (Eq. \ref{eq:deltapsi}) is calculated using the orientation of depolarisation canals ($\theta_{DC}$) as $\theta_1$ and the orientation of starlight polarisation ($\theta_\star$) as $\theta_2$. We created a sample of $\Delta \psi$ angles for each distance bin and use it to calculate PRS, its associated error and the significance threshold.

We used a distance bin size of 50 pc and a step of 10 pc, resulting in an average sample size of 18 stars per bin. This is the smallest bin size we could use, to keep the average sample size reasonable. Larger bin sizes reduce the sensitivity of this method but generally give the same result. Figure \ref{west_PRS} reveals a significant alignment between starlight polarisation and the north-south depolarisation canals in the distance range from 240 pc to 400 pc and from 450 pc to 490 pc. To determine the lower and the upper limits for the minimum distance to the Faraday structures, we searched for bins that satisfy conditions outlined in \citet{panopoulou21}. To qualify as an upper limit, a bin has to have $\Delta \psi \leq 20^\circ$ and a PRS value above the significance threshold. Both conditions must be met consistently across multiple bins. To establish the lower limit, we searched for bins where $\Delta \psi$ changes abruptly from a higher value to $\leq 20^\circ$. These conditions set the lower limit to the minimum distance at $200$ pc and the upper limit at $240$ pc. Both limits have an error of $50$ pc, as defined by the bin size. The result for the {alignment in the} north-south direction agrees with the \citet{turic21} result for the 3C196B field. However, we did not reproduce the alignment they found for the east-west depolarisation canals and stars closer than 200 pc in this, much larger field.

{When performing a tomography based on starlight polarisation measurements, we can derotate the observed polarisation of background stars for the observed polarisation of the foreground stars. In a large area, such as the western region, the foreground stars are not necessarily all tracing the same ISM and derotation has to be done on a a smaller scale. Unfortunately, the density of stars closer than 200 pc is insufficient to derotate all the background stars. The derotation was performed in the 3C196B field \citep{turic21}, where it did not significantly affect the results. }

Additionally, we examined the starlight polarisation fraction by calculating the mean polarisation fraction in the same patches, as in Sect. \ref{Sect:mag field vs lotss} (presented in Fig. \ref{pol_frac}). {Higher values for the polarisation fraction are produced by a magnetic field that is coherent along the LOS and mainly in the plane-of-sky. Stars with the highest polarisation fraction are found in the western region and in the south-eastern area of the mosaic. Although the starlight polarisation in the south-eastern region seems to have a preferred orientation (see Fig. \ref{RHT+stars}) and a high polarisation fraction, the synchrotron emission in the same area is depolarised and there are no well defined depolarisation canals, so we cannot compare the two. }  


\subsection{Depolarisation canals and HI filaments}
\label{LoTSSvsEBHIS}

\begin{figure*}
   \centering
   \includegraphics[width=\hsize]{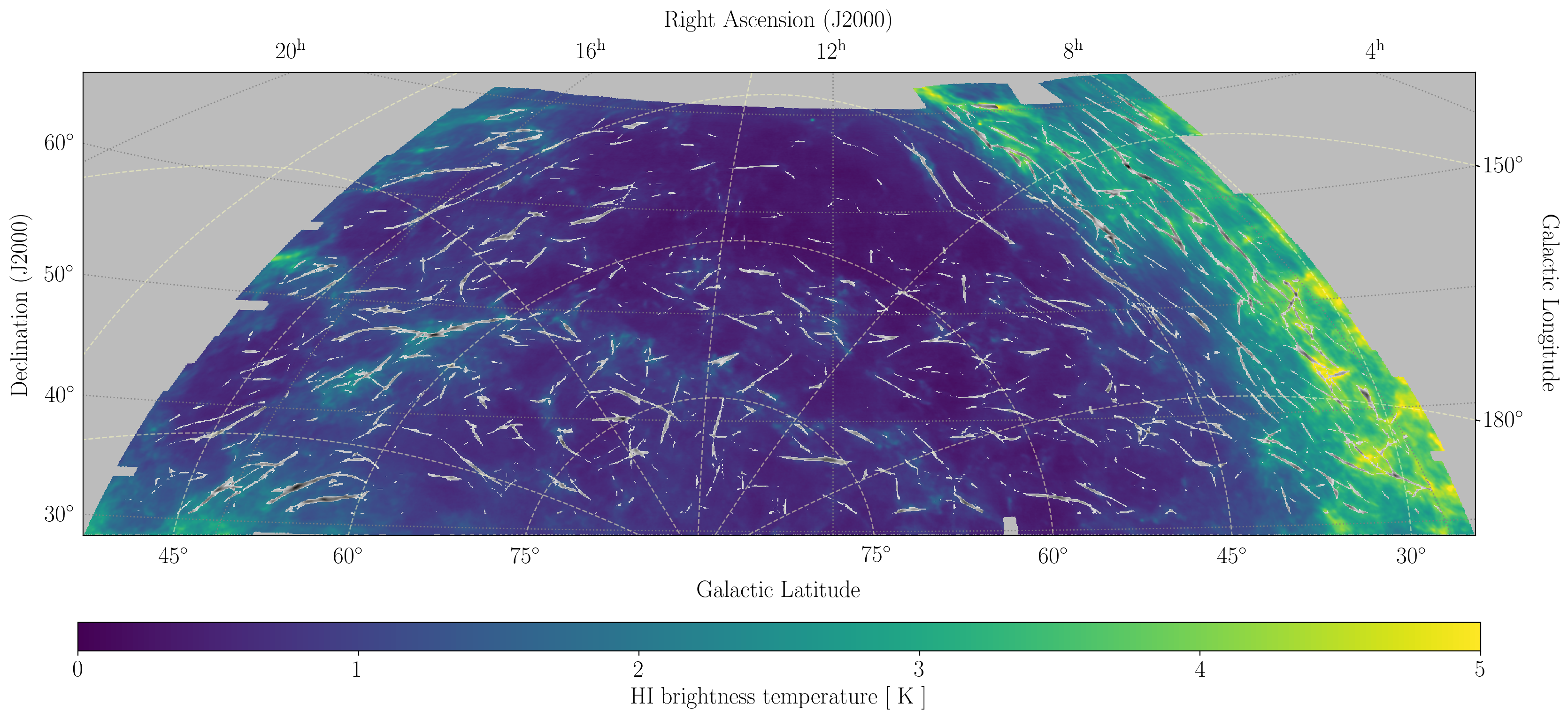}
      \caption{HI brightness temperature in the LoTSS mosaic area. Gray lines represent the linear features {of HI} detected using RHT.}   \label{HI}
\end{figure*}
 
The map of HI brightness temperature integrated over local velocities is shown in Fig. \ref{HI}. The HI emission is most prominent in the western region,  where it is organised in filaments aligned in the north-south direction, consistent with the other three tracers. We quantified the orientation of HI filaments by performing RHT on the integrated map of HI intensity using parameters ($D_K$, $D_W$, $Z$) = (10, 55, 0.7). To assess the alignment between depolarisation canals and HI filaments in the western region, we used the patch method with a patch size of 1.5 degrees (same as in Sect. \ref{Sect:mag field vs lotss}). This approach allows us to focus on alignment on larger scales, while minimising the influence of small-scale structures. The calculation confirmed the visual alignment, with a PRS value of $12.7 \pm 0.2$, well over the significance threshold of 3.1. Most alignment comes from the emission in a range from -16 to 7 $\rm{km\ s^{-1}}$, which is parallel to the Galactic disk and, at a larger scale, connected to the emission from the disk. 

\citet{clark18} and \citet{clark19} proposed that coherency of the orientation of HI filaments over a large velocity range indicates low degree of magnetic field tangling along the LOS. We used their synthetic HI Stokes parameter maps integrated over velocities to compute the synthetic HI polarisation fraction ($p_{\mathrm{HI}}$) in the western region. To estimate the uncertainty in $p_{\mathrm{HI}}$, we propagate empirical uncertainties given in \citet{clark19}. We find a result of $\langle p_{\rm{HI}} \rangle = 11.3 \pm 0.1~\%$, while the full sky average reported by the authors is $9.2~\%$. The degree of order in the plane-of-sky magnetic field in this area can be estimated using the map of polarisation angle dispersion function, $S_\mathrm{HI}$ \citep{clark19}. We find the field to be ordered, with polarisation angle dispersion of $5\degree$. These results are in line with those reported for the 3C196 field \citep{clark19} and for the 3C196 B field \citep{turic21}. High $p_{\rm{HI}}$ points to a coherent magnetic field along the LOS, which agrees well with the conclusions we drew from the starlight polarisation fraction.

Additionally, we conducted a visual inspection of the entire HI cube and the Faraday cube. Still, we did not find other obvious morphological correlation between HI filaments and depolarisation canals or diffuse emission for any velocity channel-and-Faraday depth combination.

\section{Discussion}
\label{sec:discussion}

\begin{figure*}
   \centering
   \includegraphics[width=\hsize]{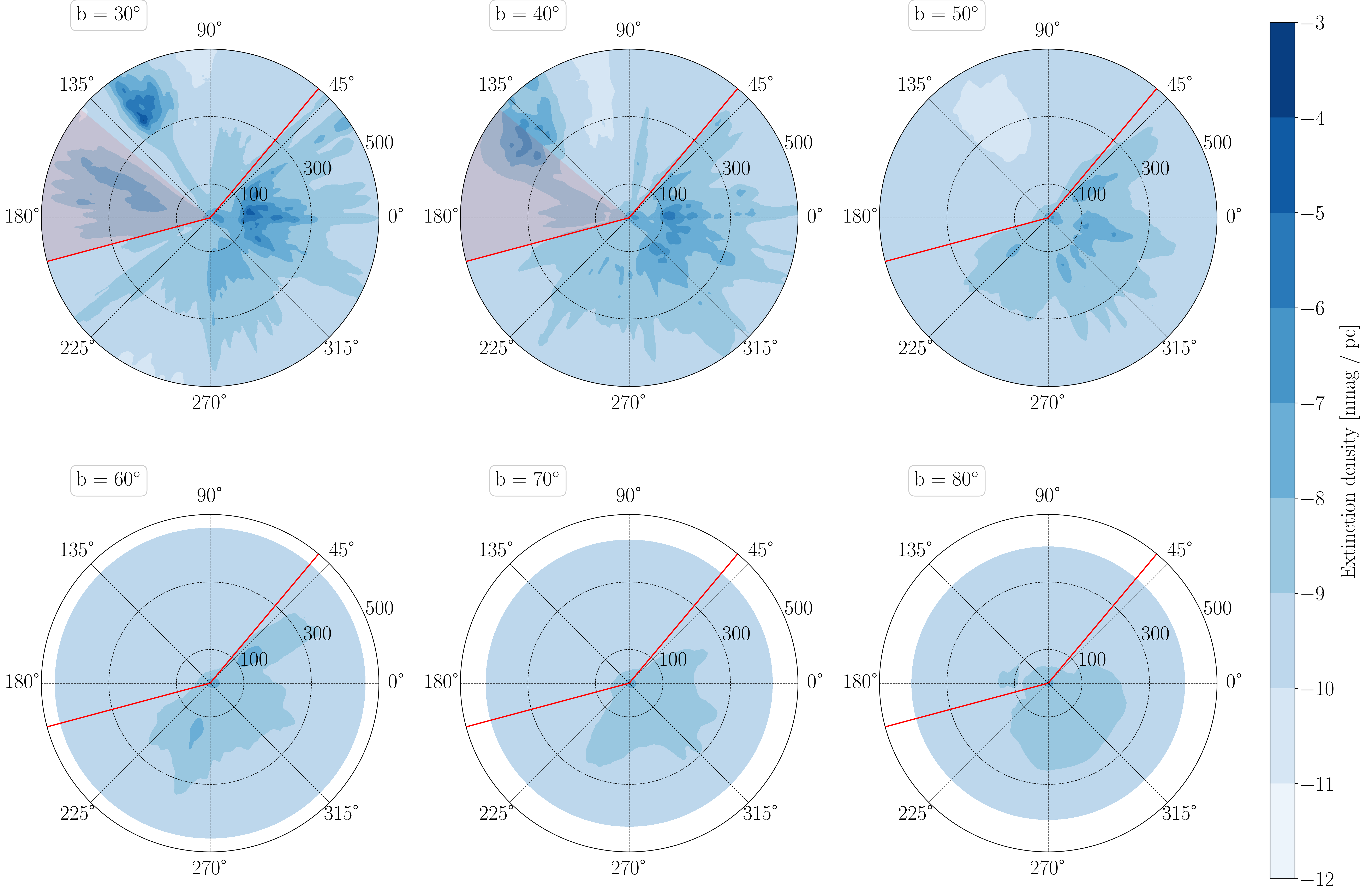}
      \caption{Each panel in this figure represents a dust extinction map created using \citet{vergely22} data. The maps are plotted in Galactic coordinates, with the Sun in the centre. The angular coordinate denotes different Galactic longitudes, and the radial coordinate represents the distance from the Sun. Each panel is at a different Galactic latitude. Red lines represent borders of the LoTSS mosaic, and red shaded areas mark the location of the western region.}   \label{Vergely_maps}
\end{figure*}

In the previous section, we established distance constraints for the structures observed in the western region. In this section, we explore the potential origins of these structures.

One possible source of the structures associated with observed depolarisation canals is the wall of the Local Bubble. \citet{pelgrims20} modelled the geometry of the Local Bubble, using dust extinction maps from \citet{lallement19} to reconstruct the shape of the Local Bubble wall. We used the full-sky map of distances to the Local Bubble \citep{pelgrims20}, which was constructed by expanding the Local Bubble shell into spherical harmonic expansion up to multipoles $l_{max}$ of 10. The median distance and the interquartile range of the distribution of distances to the Local Bubble inner wall in the western region are 254 pc and 24 pc, respectively. These values agree well with the estimated minimum distance towards the structures associated with depolarisation canals. The \citet{pelgrims20} model can also be used to determine whether the surface of the Local Bubble in a specific direction is on the plane of sky or along the LOS. Suppose we assume that the magnetic field is tangential to the surface of the Local Bubble. In that case, we can tie the surface orientation to the orientation of the magnetic field within the Local Bubble wall. Our calculation indicates that in the western region, the Local Bubble's wall is primarily on the plane of the sky. A plane-of-sky magnetic field (supported by the results in Sects. \ref{stellar pol vs dep canals} and \ref{LoTSSvsEBHIS}) could produce the observed alignment between the HI, starlight polarisation, and depolarisation canals. Therefore, we conclude that the magnetic field on the surface of the Local Bubble is a good candidate for the origin of Faraday structures and associated depolarisation canals in the western region and for the alignment they exhibit with other tracers. 

The alignment of depolarisation canals, HI filaments, and the magnetic field in the western region is almost parallel to the Galactic plane, which prompts us to examine two other local structures in the outer Galaxy: the Radcliffe wave \citep{alves20} and the Local Arm \citep{reid16}. Although they are parallel to the Galactic plane, at the Galactic longitudes of the western region ($150^\circ$ to $190^\circ$), both the Radcliffe wave and the Local Arm lie below the Galactic plane,  making it unlikely for the structures to be connected with depolarisation canals we observe.

In the remaining areas of the mosaic, we observed no alignment of depolarisation canals with either the magnetic field or HI filaments. To better understand the difference between the area showing a strong alignment and the rest of the mosaic, we used the dust extinction maps of \citet{vergely22}. These maps are an upgrade to the \citet{lallement19, lallement22} maps because they are created through successful cross-calibration of photometric and spectrophotometric information. This makes the maps more detailed and better at mapping faint high-latitude clouds, making them better suited to our purposes. We used the dust extinction data cube with a resolution of $10~\rm{pc}$, covering an area of $3000~\rm{pc} \times 3000~\rm{pc} \times 800~\rm{pc}$ to plot six dust maps in Fig. \ref{Vergely_maps}. The plots are presented in Galactic coordinates, with the centre of each plot located at the Sun. The radial lines represent LOSs at different Galactic longitudes, while the Galactic latitude b is constant for each panel. As the latitude increases, LOS becomes shorter due to the shape of the dust extinction data cube. However, the specific areas of interest remain visible in all plots. 

The LoTSS mosaic covers the Galactic latitude range from $\sim30^\circ$ to $\sim75^\circ$ and the Galactic longitude range from roughly $50^\circ$ to $195^\circ$. Boundaries of the mosaic in longitude are indicated with two red lines in all the plots of Fig. \ref{Vergely_maps}. The western region, situated between Galactic longitudes $140^\circ$ and $195^\circ$ and Galactic latitudes $30^\circ$ and $45^\circ$, is visible in the first two plots, where it is shaded in red. In this region, the areas of increased dust density appear further away with increasing latitude, between $100$ and $300~\rm{pc}$, in line with the distance limits we established in Sect. \ref{stellar pol vs dep canals}. Another area of increased intensity is visible in the first two panels, at latitudes $30^\circ$ and $40^\circ$, just beyond the western region at longitudes $140^\circ \lesssim l \lesssim 110^\circ$. When viewed in equatorial coordinates (e.g. Fig. \ref{RHT+stars}), this area is located at declinations above $70^\circ$ (above Loop III) and falls outside of the coverage of the LoTSS mosaic. 

Apart from the two areas mentioned above, no other local regions of high dust extinction exist within the LoTSS mosaic. This indicates that it is likely that in the area beyond the western region, \textit{Planck}'s plane-of-sky magnetic field is tracing dust structures located further away than 500 pc. Additionally, the HI temperature brightness in other parts of the mosaic is lower compared to the western region (see Fig. \ref{HI}). The lack of a sharp jump in the density of the neutral medium indicates the presence of a hole in the wall of the Local Bubble in these directions. Furthermore, LoTSS emission away from the western region and the loop is faint and patchy, without any well-defined depolarisation canals. This can be caused by low intrinsic polarised emission along the LOS in that direction or depth depolarisation along the LOS. We conclude that most of the polarised emission observed by  LOFAR in this area can be associated with local structures. 

\section{Conclusions}
\label{sec:conclusions}

In this paper, we build on the work of E22 by conducting a multi-tracer analysis of the LoTSS mosaic. We searched for areas where depolarisation canals align with the magnetic field lines and starlight polarisation. We found one such location, namely, the western region, covering an area between Dec of $48^\circ$ and $70^\circ$, and RA of $8\rm{^h}10\rm{^m}$ to $14\rm{^h}0\rm{^m}$. We followed the method of \citet{panopoulou21} and used PRS to find the distance at which the starlight polarisation aligns with the depolarisation canals. By doing so, we have set the lower and the upper limit to the minimum distance to the structures associated with depolarisation canals in the western region to 200 and 240 pc, respectively. The estimated minimum distance corresponds to the distances at which the \citet{pelgrims20} map puts the inner edge of the Local Bubble wall along these lines of sight.

Additionally, we found an alignment between HI filaments and depolarisation canals in the same region. While our results confirm the alignment found by \citet{jelic18} and \citet{turic21} on a much larger area, they also demonstrate that this alignment between the three tracers is not universal, as it is not present in the larger part of the mosaic. By inspecting the dust extinction maps from \citet{vergely22}, we concluded that the absence of any alignment coincides with low HI brightness temperature and low levels of dust extinction, suggesting a lack of a distinct boundary for the Local Bubble wall in those areas. This finding confirms that LOFAR is mainly sensitive to the local synchrotron emitting and Faraday rotating structures. 

\begin{acknowledgements}
We thank the anonymous referee for their comments that helped improve this paper. A.E., V.J. and L.T. acknowledge support from the Croatian Science Foundation for project IP-2018-01-2889 (LowFreqCRO). M.H. acknowledges funding from the European Research Council (ERC) under the European Union's Horizon 2020 research and innovation programme (grant agreement No 772663). A.B. acknowledges support from the INAF initiative "IAF Astronomy Fellowships in Italy" (grant name MEGASKAT). 
This paper is based on data obtained with the International LOFAR Telescope (ILT) under project codes LC2$\_$038 and LC3$\_$008. LOFAR (van Haarlem et al. 2013) is the Low Frequency Array designed and constructed by ASTRON. It has observing, data processing, and data storage facilities in several countries owned by various parties (each with its own funding sources) and collectively operated by the ILT foundation under a joint scientific policy. The ILT resources have benefitted from the following recent major funding sources: CNRS-INSU, Observatoire de Paris and Université d’Orléans, France; BMBF, MIWF-NRW, MPG, Germany; Science Foundation Ireland (SFI), Department of Business, Enterprise and Innovation (DBEI), Ireland; NWO, The Netherlands; The Science and Technology Facilities Council, UK; Ministry of Science and Higher Education, Poland.
\end{acknowledgements}

\bibliographystyle{aa}
\bibliography{reference_list.bib} 


\end{document}